\documentclass[aps,pre,twocolumn,superscriptaddress,amssymb]{revtex4-2}
\usepackage{physics}
\usepackage[dvipdfmx]{graphicx}
\usepackage{xcolor}
\begin{document}

\title{Tensor-Network Population Annealing}


\author{Takumi Oshima}
\affiliation{
 Graduate School of Arts and Sciences,
 The University of Tokyo,
 Komaba, Meguro-ku, Tokyo 153-8902, Japan
}
\author{Yuma Ichikawa}
\email{ichikawa.yuma@fujitsu.com}
\affiliation{
 Fujitsu Limited,
 4-1-1 Kamikodanaka, Nakahara-ku, Kawasaki-shi, Kanagawa 211-8588, Japan
}
\affiliation{
 RIKEN Center for Advanced Intelligence Project,
 Nihonbashi 1-chome Mitsui Building, 15th floor,
 1-4-1 Nihonbashi, Chuo-ku, Tokyo 103-0027, Japan
}
\author{Koji Hukushima}
\email{k-hukushima@g.ecc.u-tokyo.ac.jp}
\affiliation{
 Graduate School of Arts and Sciences,
 The University of Tokyo,
 Komaba, Meguro-ku, Tokyo 153-8902, Japan
}
\affiliation{
Komaba Institute for Science, The
University of Tokyo, 3-8-1 Komaba, Meguro-ku, Tokyo 153-8902, Japan
}


\date{\today}

\begin{abstract}
We propose a hybrid sampling method, tensor-network population annealing (TNPA), which combines tensor-network (TN) initialization with population annealing (PA). We apply this method to the two-dimensional $\pm J$ Edwards-Anderson Ising spin glass. The approach is motivated by the limitations of existing methods: TN-based samplers can become numerically unstable in frustrated spin systems at low temperatures, whereas conventional PA requires a long annealing schedule when started from the high-temperature limit. In TNPA, TN contractions are used only within a reliable temperature range to generate initial configurations that are close to the equilibrium distribution. The subsequent low-temperature equilibration is then carried out by PA. To stabilize the initialization process, we introduce a diagnostic based on the effective sample size that adaptively selects the initialization temperature. The proposed framework provides a practical and physically motivated route to low-temperature sampling by combining the complementary strengths of TN and PA.  
\end{abstract}


\maketitle
\section{Introduction}
\label{sec:introduction}

One of the central challenges in many-body physics is the evaluation of high-dimensional sums and integrals, as required, for example, in the computation of partition functions and thermodynamic observables. A standard numerical approach to this difficulty is stochastic sampling, particularly through Markov-chain Monte Carlo (MCMC) methods~\cite{Landau_Binder_2014, NewmanBarkema1}. Although MCMC methods have been successfully applied to a wide range of problems in statistical physics and related fields, the slow relaxation of the Markov chain remains a serious obstacle in systems with complex free-energy landscapes, such as spin glasses~\cite{BinderYoung, spin_glass_beyond} and polymers~\cite{degennes_scaling}. To overcome this difficulty, a variety of advanced MCMC methods, including cluster-update algorithms~\cite{SwendsenWang_1987, Wolff_1989, WANG1990} and extended-ensemble methods~\cite{Iba_2001}, have been developed. Among the latter, the exchange Monte Carlo method (EXMC)~\cite{Geyer1991,HukushimaNemoto}, also known as parallel tempering~\cite{Marinari1998}, is widely used in spin-glass simulations. By simulating replicas at multiple temperatures and exchanging configurations between neighboring temperatures, EXMC allows low-temperature replicas to overcome free-energy barriers by transitioning to higher temperatures. 

Meanwhile, tensor-network (TN) methods have emerged as a powerful framework for approximating high-dimensional sums~\cite{ORUS2014117,TN_review}. In classical statistical-mechanics models, local Boltzmann weights can be represented as tensors, so that the partition function can be expressed as a tensor network. From this viewpoint, several approximation and coarse-graining methods have been developed, including the tensor renormalization group~\cite{TRG} and its extensions~\cite{Xie_2009, Xie_2012, TNR}, the corner transfer-matrix renormalization group~\cite{CTMRG}, and transfer-matrix approaches based on time-evolving block decimation~\cite{Vidal_2004, OrusVidal_2008}. 

These two frameworks are complementary: MCMC methods provide asymptotically exact sampling but may suffer from slow relaxation, whereas TN methods enable direct approximation of the partition function but can become inaccurate in strongly frustrated or low-temperature regimes.  
Recent developments in computational physics have explored hybrid approaches that integrate TN methods with Monte Carlo (MC) sampling. These approaches can be broadly categorized into two groups. One approach is to integrate stochastic sampling within TN methods, for instance, by employing MC updates to evaluate high-dimensional TN contractions~\cite{ferris2015,huggins2017,todo2024}. The other uses TN as a surrogate model to guide MC sampling, for example, by constructing an approximate proposal distribution~\cite{FriasPerez_2023, Chen_2025, chen2025_3d}.  

In this work, we follow the latter approach and use TN methods to estimate the conditional probabilities of the canonical distribution. This enables the sequential generation of configurations from a TN-based approximation. 
When these conditional probabilities are evaluated with sufficient accuracy, this framework can also enable direct independent sampling~\cite{Oshima2026}. 
More generally, however, the conditional probabilities are only approximately evaluated, and practical sampling requires a correction step. This idea forms the basis of the tensor-network Monte Carlo method (TNMC)~\cite{Chen_2025, Chen202509, chen2025_3d}, in which TN-based proposals are combined with a Metropolis-Hastings filter~\cite{Metropolis_1953, Hastings_1970} to ensure convergence to the target distribution.  

This framework, however, has a serious limitation in frustrated spin systems. The difficulty lies in the TN-based evaluation of the conditional probabilities, which can become unstable at low temperatures. This issue is particularly pronounced in spin-glass models, where competing interactions lead to highly nontrivial correlations and cancellations. At low temperatures, TN contraction schemes based on boundary matrix-product-state representations can become ill-conditioned, leading to numerical instabilities and inaccurate estimates of conditional probabilities. It has been pointed out that this may even produce negative estimates of partition functions~\cite{Wang_2014} or require calculations with extremely high numerical precision~\cite{zhu2019}. 

\begin{figure}[t]
   \centering
   \includegraphics[width=\linewidth]{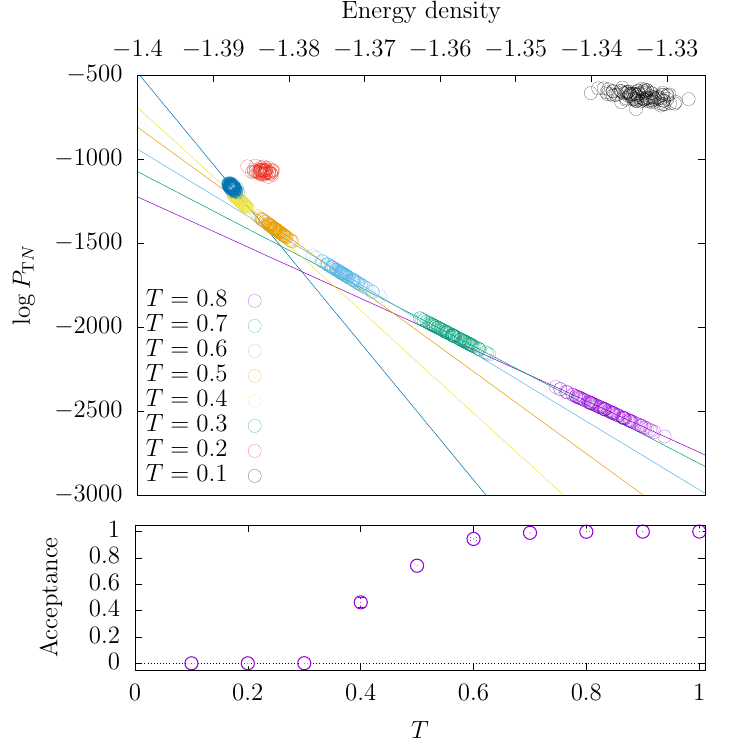}
   \caption{Top: TN-estimated log-probability, $\log P_{\mathrm{TN}}$, plotted against the energy density for configurations generated by the TN-based sampler for a typical disorder instance of the two-dimensional Edwards-Anderson Ising spin-glass model with $N=128^2$ spins. Here, $P_{\mathrm{TN}}$ is computed as the product of the approximate conditional probabilities obtained during autoregressive generation. The precise model definition and units are given later. The straight lines represent the linear relation expected for the canonical distribution at each temperature. Bottom: Temperature dependence of the Metropolis--Hastings acceptance probability in TNMC for the same disorder instance.  
   }
   \label{fig:logP}
 \end{figure}

This breakdown is illustrated in Fig.~\ref{fig:logP}. The top panel shows the TN-estimated log-probability $\log P_{\mathrm{TN}}$ of generated configurations plotted against their energy density for a typical two-dimensional spin-glass instance of size $N=128^2$, whereas the bottom panel shows the corresponding Metropolis-Hastings acceptance probability in TNMC as a function of temperature. For an exact canonical distribution, the log-probability and the energy density should be linearly related at each temperature. In the present data, the samples at relatively high temperatures follow this expected linear trend reasonably well, whereas at lower temperatures, particularly $T=0.2$ and $0.1$, they deviate markedly from it. This indicates that the TN-based sampler no longer provides a faithful approximation to the canonical distribution in the low-temperature regime. Consequently, the lower panel shows that the Metropolis-Hastings acceptance probability remains high at moderate temperatures but drops rapidly once the deviation becomes pronounced, reaching nearly zero in the low-temperature regime. As a result, TNMC can no longer make meaningful sampling progress. 

 While this limitation emphasizes the difficulty of using TN-based sampling directly at low temperatures, it also raises a complementary issue from the perspective of population-based MC methods~\cite{Iba_PMC}. Population annealing (PA)~\cite{HukushimaIba}, a population-based MC scheme successfully applied to spin-glass models~\cite{Wang_2015, Barzegar_2018}, evolves a population of replicas toward low-temperature equilibrium distributions by gradually annealing it from a high-temperature distribution. However, PA can also become less efficient at low temperatures. In systems with complex free-energy landscapes, slow relaxation can cause replicas in a finite population to become increasingly correlated, resulting in reducing population diversity. Moreover, when the target temperature is far below the high-temperature limit, PA requires a long annealing schedule with many intermediate temperatures, which increases the computational cost and may further reduce population diversity. 

 These observations suggest that neither TNMC nor conventional PA is fully satisfactory on its own. TNMC becomes unstable at low temperatures, whereas conventional PA can lose population diversity along the long annealing path from the high-temperature limit. To overcome these limitations, we propose a hybrid approach, tensor-network population annealing (TNPA), which uses TN-based sampling at an intermediate temperature where the TN contraction remains reliable and then employ PA to anneal the resulting population toward lower temperatures. By replacing the high-temperature initialization of conventional PA with TN-based initialization at a finite temperature, TNPA shortens the annealing path and allows PA to enter the difficult low-temperature regime with a more diverse population. 

 The rest of this paper is organized as follows. In Sec.~\ref{sec:methods}, following the setup, we describe the proposed TNPA algorithm, conventional PA, and the procedure for bridging TN-based sampling to the canonical distribution. We also describe the effective-sample-size-based diagnostics, outlier-removal procedure, and the adaptive selection of the initialization temperature, which are technical yet crucial components of TNPA. Sec.~\ref{sec:results} presents numerical results for the two-dimensional $\pm J$ Ising spin glass, including the implementation details of TNPA, single-instance benchmarks against existing methods, disorder-averaged entropy estimates, and an estimate of the residual entropy. Sec.~\ref{sec:summary} concludes the paper with a summary of the main results and a discussion of their implications. 
 Additional implementation details concerning the combination of ICM with PA, the comparison of resampling schemes, and the adaptive annealing schedules used in TNPA are provided in Appendices~\ref{sec:combinationICMPA}, \ref{sec:resampling_family_entropy} and \ref{sec:temperature_schedule}, respectively. 
 
\section{Methods}
\label{sec:methods}

\subsection{Canonical distribution and autoregressive factorization}
We consider a classical spin system defined on a lattice with $N$ sites. A microscopic configuration of the system is denoted by $\vb*{\sigma}=(\sigma_1,\cdots,\sigma_N)$, where each spin variable $\sigma_i$ takes a discrete value. For a Hamiltonian $\mathcal{H}(\vb*{\sigma})$, the equilibrium distribution at inverse temperature $\beta$ is given by the canonical distribution
\begin{equation}
\label{eqn:canonical}
P_{\mathrm{C}}(\vb*{\sigma}) = \frac{1}{Z(\beta)}\exp\left(-\beta \mathcal{H}(\vb*{\sigma})\right) = \frac{1}{Z(\beta)}\tilde{P}_{\mathrm{C}}(\vb*{\sigma}),
\end{equation}
where $Z(\beta)$ is the partition function, and $\tilde{P}_\mathrm{C}(\vb{\sigma})$ denotes the corresponding unnormalized weight. 

In the following, we assume a fixed ordering of the spins. For any such ordering, the canonical distribution has an exact autoregressive factorization according to the chain rule of probability:  
\begin{equation}
\label{eqn:conditional}
P_{\mathrm{C}}(\vb*{\sigma}) = P_{\mathrm{C}}(\sigma_1) \prod_{i=2}^N P_{\mathrm{C}}(\sigma_i | \vb*{\sigma}_{<i}),
\end{equation}
where $\vb*{\sigma}_{<i} = (\sigma_1, \cdots, \sigma_{i-1})$. For a fixed ordering, the conditional probability of spin $i$ given $\vb*\sigma_{<i}$ can be expressed as  
\begin{equation}
\label{eqn:conditional_partition}
    P_{\mathrm{C}}(\sigma_i | \vb*{\sigma}_{<i}) = \frac{\sum_{\vb*{\sigma}_{>i}} \exp\left(-\beta \mathcal{H}(\vb*{\sigma})\right)}{\sum_{\vb*{\sigma}_{>i-1}} \exp\left(-\beta \mathcal{H}(\vb*{\sigma})\right)} = \frac{Z(\sigma_i, \vb*{\sigma}_{<i})}{\sum_{\sigma_i}Z(\sigma_i, \vb*{\sigma}_{<i})}.
\end{equation}
where $\vb*{\sigma}_{>i} = (\sigma_{i+1}, \cdots, \sigma_N)$, and $Z(\sigma_i,\vb*{\sigma}_{<i})$ denotes the conditional partition function obtained by summing over the remaining spins $\vb*{\sigma}_{>i}$. This representation forms the basis for sequential sampling. In fact, it has already been used in autoregressive sampling approaches, including neural-network-based methods~\cite{Wu_2019, McNaughton_2020, Bono_PRE_2025, Del_Bono_2025} and TN-based sampling methods such as TNMC~\cite{FriasPerez_2023, Chen_2025, Chen202509, chen2025_3d}. However, evaluating these conditional probabilities exactly is intractable for large systems, which motivates the use of TN approximations. 

\subsection{Tensor-Network-based sampling}
Since an exact evaluation of the conditional partition functions in Eq.~(\ref{eqn:conditional_partition}) is generally intractable for large systems, we approximate them using tensor-network contractions. This approximation defines the TN-based autoregressive distribution used to initialize TNPA. The local Boltzmann weights associated with the interactions are factorized into local tensors whose auxiliary indices are contracted along the bonds of the lattice. Summing over the physical spin variables yields a TN representation of the full partition function. Similarly, the conditional partition function $Z(\sigma_i,\vb*{\sigma}_{<i})$ is represented by fixing the specified physical spin variables $\{\sigma_i,\vb*{\sigma}_{<i}\}$ in the corresponding local tensors while summing over all the remaining spin variables. 

The resulting TN is contracted using a boundary matrix-product-state (MPS) scheme based on time-evolving block decimation (TEBD)~\cite{Vidal_2004, OrusVidal_2008}. In this scheme, each successive row of local tensors is represented as a transfer matrix-product operator acting on the boundary MPS. After each row is incorporated MPS is compressed by singular-value decomposition so that its bond dimension does not exceed a prescribed value $\chi$. The bond dimension $\chi$ therefore controls the trade-off between the accuracy and computational cost of the contraction. Unlike tensor-renormalization-group methods, this procedure contracts the TN sequentially from one boundary and does not involve a real-space coarse-graining of the two-dimensional lattice. 

Because the finite $\chi$ contraction provides only an approximation, we denote the resulting estimate of the conditional partition function by $\widehat{Z}_\chi(\sigma_i,\vb*\sigma_{<i})$. The conditional probability used for TN-based sampling is then given by 
\begin{equation}
    P_{\mathrm{TN}}(\sigma_i\mid \vb*\sigma_{<i})=\frac{\widehat{Z}_\chi(\sigma_i,\vb*\sigma_{<i})}{\sum_{\sigma_i'}\widehat{Z}_\chi(\sigma_i',\vb*\sigma_{<i})}.
\end{equation}
By sequentially sampling each spin from these conditional probabilities, we obtain the autoregressive TN distribution
\begin{equation}
    P_{\mathrm{TN}}(\vb*\sigma)= P_{\mathrm{TN}}(\sigma_1)\prod_{i=2}^N P_{\mathrm{TN}}(\sigma_i\mid\vb*\sigma_{<i}).
    \label{eqn:ptn_autoregressive}
\end{equation}
When the conditional partition functions are evaluated exactly, this product coincides with the canonical distribution in Eq.~(\ref{eqn:canonical}) for any spin ordering. For a finite bond dimension $\chi$, however, the approximate contractions do not generally preserve this exact relation, and the resulting distribution $P_{\mathrm{TN}}$ may depend on the chosen ordering because the accumulated truncation errors depend on the contraction path. Accordingly, the value of $P_{\mathrm{TN}}$ assigned to each generated configuration is evaluated directly as the product of the approximate conditional probabilities obtained during sequential sampling. 

\subsection{Tensor Network Population Annealing (TNPA)}
\label{sec:tnpa}

As discussed in Sec.~\ref{sec:introduction}, TNMC becomes ineffective at low temperature in frustrated spin systems, most notably because of the collapse of its acceptance rate. To address this limitation, we propose TNPA, which combines TN-based initialization with subsequent population annealing. The key idea is to use TN-based autoregressive distribution $P_{\mathrm{TN}}$, defined in the previous subsection, to generate an initial population at a temperature chosen within the regime where the TN contraction remains numerically reliable. Because the finite-$\chi$ distribution $P_{\mathrm{TN}}$ only approximates the canonical distribution, the discrepancy between the two distributions is accounted for through importance reweighting at the initialization temperature. Thereafter, the method reduces to standard PA: the population of replicas is annealed to the target temperature using weight updates, resampling, and equilibration sweeps.

Conceptually, TNPA assigns distinct and complementary roles to TN and PA. TN-based autoregressive sampling can independently generate configurations at an intermediate temperature where the TN contraction remains reliable, but its direct application becomes unreliable at lower temperatures because of numerical instabilities.  TNPA therefore uses TN as an initializer rather than as a low-temperature sampler. From the perspective of PA, this finite-temperature initialization provides a diverse population closer to the target temperature and shortens the annealing path from the high-temperature limit. PA then carries this population toward lower temperatures, beyond the range where TN-based sampling can be applied reliably. This complementarity does not imply that PA is intrinsically efficient at arbitrarily low temperatures. Rather, TN initialization helps PA enter the difficult low-temperature regime with less accumulated loss of population diversity, while PA extends the TN-generated population beyond the reliable temperature range of the TN contraction.

The essential ingredient of TNPA is therefore not a modification of the PA procedure itself, but a TN-guided initialization together with a stabilization procedure based on reweighting, effective-sample-size (ESS) diagnostics, outlier removal, and adaptive selection of the initialization temperature. In what follows, we first recall the PA framework, then describe the TN-guided initialization and its reweighting correction, and present the full TNPA workflow and the associated implementation details. 

\subsection{Population Annealing (PA)}
\label{sec:pa}

PA~\cite{HukushimaIba} is a population-based Monte Carlo scheme~\cite{Iba_PMC} in which a
population of replicas is progressively cooled along an inverse‐temperature
schedule $\beta_0 < \beta_1 < \cdots < \beta_n$. At each temperature step, the replica weights are updated, the population is resampled to control weight degeneracy, and short MCMC sweeps are performed to maintain population diversity. In this way, PA evolves the population of replicas toward the canonical distribution at each temperature.

In the standard setting, the initialization is performed at $\beta_0=0$, where sampling is straightforward for spin systems. For Ising spins, this means that each spin independently takes the value $\pm1$ with probability $1/2$. Thus, $R$ initial replicas $\{\vb*{\sigma}_k^{(0)}\}_{k=1}^R$ are drawn independently from the uniform distribution and assigned unit weight, $W_k^{(0)}=1$. Let $\vb*{\sigma}_k^{(i)}$ denote the $k$th replica at $\beta_i$ with energy $E_k^{(i)}=\mathcal{H}(\vb*{\sigma}_k^{(i)})$. 

When the inverse temperature is updated from $\beta_i$ to $\beta_{i+1}$, the weights are also updated according to 
 \begin{equation}
    W_k^{(i+1)} \;=\; W_k^{(i)} \, e^{-(\beta_{i+1}-\beta_i)\,E_k^{(i)}} .
    \label{eq:pa-weight}
\end{equation}
After normalization, the replicas are resampled with replacement, and the resampled population is again assigned a unit weight. Short MCMC sweeps are then performed at $\beta_{i+1}$ to restore population diversity. 

The efficiency of PA is mainly governed by the population size $R$, the number of equilibration sweeps at each temperature, and the choice of temperature increments $\Delta\beta_i=\beta_{i+1}-\beta_i$, which must be chosen so that weight degeneracy remains under control.  
In standard PA, the initialization is thus fixed to the trivial high-temperature limit $\beta_0=0$. In the following, we replace this initialization with a TN-based procedure while preserving the PA framework.

\subsection{Single-step bridge from $P_{\mathrm{TN}}$ to the canonical distribution}
\label{sec:single-bridge}

In the PA scheme described above, sampling from the canonical distribution is achieved by bridging a source distribution through an annealing process. In most applications, the high-temperature limit $\beta_0=0$ is used as the source distribution since sampling is straightforward. However, when the temperature of interest is not sufficiently close to $\beta=0$, a long and carefully tuned annealing schedule may be required. 

In the TNPA scheme proposed here, we instead use a TN-based approximation as the source distribution. Specifically, we consider the autoregressive distribution $P_{\mathrm{TN}}(\vb*{\sigma})$, which approximates the canonical distribution at an inverse temperature $\beta_0>0$. Starting from $P_\mathrm{TN}$ is carried out sequentially, with the required conditional probabilities approximated by TN contractions of the corresponding conditional partition functions. In practice, these contractions are performed with a finite bond dimension $\chi$, introducing a controlled approximation through tensor truncation. 

To connect this approximate source distribution to the canonical distribution at $\beta_0$, we introduce the interpolation family 
\begin{equation}
  P_{\mathrm{int}}(\vb*{\sigma};\alpha)\;\propto\;
  \big(P_{\mathrm{TN}}(\vb*{\sigma})\big)^{\alpha}
  \big(\tilde P_{\mathrm{C}}(\vb*{\sigma})\big)^{1-\alpha},
  \qquad \alpha\in[0,1]
  \label{eq:interp}
\end{equation}
which continuously interpolates between $P_{\mathrm{TN}}$ at $\alpha=1$ and $\tilde{P}_\mathrm{C}$ at $\alpha=0$. 

In principle, PA could be applied over a sequence of intermediate $\alpha$ values. However, this would require TN contractions at each $\alpha$ in order to perform local MCMC sweeps, which is computationally impractical. We therefore adopt a single-step bridge, yielding a one-shot importance reweighting from $P_{\mathrm{TN}}$ to the canonical distribution at $\beta_0$. 

Concretely, we draw $R$ replicas $\{\vb*{\sigma}_k^{(0)}\}_{k=1}^{R}\!\sim\!P_{\mathrm{TN}}$ and assign initial weights
\begin{equation}
  W_k^{(0)} \;=\;
  \frac{\tilde P_{\mathrm{C}}(\vb*{\sigma}_k)}{P_{\mathrm{TN}}(\vb*{\sigma}_k)}
  \;=\;
  \frac{e^{-\beta_0\,\mathcal{H}(\vb*{\sigma}_k)}}
       {P_{\mathrm{TN}}(\vb*{\sigma}_k)} ,
  \qquad k=1,\dots,R.
  \label{eq:single-bridge-weight}
\end{equation}
This corresponds to an importance-sampling correction of the TN approximation at the initial temperature. After resampling according to $W_k^{(0)}$ and resetting all weights to unity, the resulting population approximates the canonical distribution at $\beta_0$ without requiring additional TN contractions.

\subsection{ESS-based diagnosis and outlier removal}
\label{sec:ess-diagnostic}

Given the TN-generated replicas $\{\vb*{\sigma}_k^{(0)}\}_{k=1}^R$ at
the initialization inverse temperature $\beta_0$, we consider the single–step bridge weights defined by Eq.~\eqref{eq:single-bridge-weight}. The effective sample size (ESS) computed from these weights is defined as 
\begin{equation}
  \mathrm{ESS}
  \;=\;
  \frac{\left(\sum_{k=1}^{R} W_k^{(0)}\right)^2}
       {\sum_{k=1}^{R} \left(W_k^{(0)}\right)^2}
  \;\in\;[1,\,R] ,
  \label{eq:ess-noalpha}
\end{equation}
and provides a standard measure of the overlap between the TN-based sampling distribution $P_{\mathrm{TN}}$ and the canonical distribution $P_\mathrm{C}$ at $\beta_0$. 

Under standard importance-sampling assumptions, the ESS has a well-known interpretation in terms of the $\chi^2$-divergence between two distributions. Defining   
\begin{equation}
    \chi^2(P_\mathrm{C}\,\|\,P_\mathrm{TN}) = \sum_{\vb*{\sigma}}P_\mathrm{TN}(\vb*\sigma)\left(\frac{P_\mathrm{C}(\vb*\sigma)}{P_\mathrm{TN}(\vb*\sigma)}-1\right)^2,
\end{equation}
we obtain, in expectation, 
\begin{equation}
    \frac{\mathrm{ESS}}{R} \approx \frac{1}{1+\chi^2(P_\mathrm{C}\,\|\, P_\mathrm{TN})}.
\end{equation}
Thus, a small $\mathrm{ESS}$ corresponds to a large $\chi^2$-divergence, i.e., poor overlap between $P_\mathrm{TN}$ and $P_\mathrm{C}$ at $\beta_0$. If $P_\mathrm{TN}$ is close to $P_\mathrm{C}$, the weights $W_k^{(0)}$ are nearly uniform and $\mathrm{ESS}\approx R$. Conversely, poor overlap yields highly variable weights, with a few samples dominating and significantly reducing $\mathrm{ESS}$, indicating that the TN-target discrepancy is too large at the chosen initial temperature. 

\begin{figure}
    \centering
    \includegraphics[width=\linewidth]{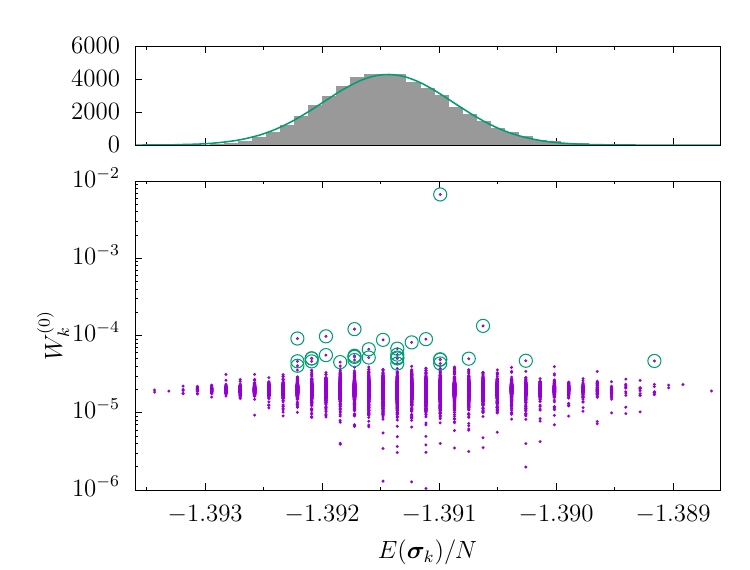}
    \caption{Top: Energy-density histogram for $R=10^5$ configurations generated by the tensor-network (TN) approximation for a two-dimensional Edwards--Anderson $\pm J$ Ising spin-glass model with $L=128$ at temperature $T=0.4$. The precise model definition and units are given later. Bottom: Scatter plot of the corresponding relative weights versus energy density. A small number of configurations carry disproportionately large weights, leading to a significant reduction of the effective sample size (ESS). By discarding the outliers denoted by the circled points, the ESS ratio increases from 0.31 to 0.99. }
    \label{fig:reweight_tn}
\end{figure}

In standard importance sampling or PA, configurations with large weights typically correspond to regions of high probability under the target distribution, and resampling appropriately allocates more replicas to such configurations. In the TN setting, however, large weights may arise from inaccuracies in the TN approximation rather than from genuine features of the target distribution. As a result, a small number of configurations can acquire anomalously large weights that dominate the resampling step. This leads to a pathological situation in which resampling amplifies approximation errors rather than correcting them. 
This behavior is illustrated in Fig.~\ref{fig:reweight_tn}. Although the energy-density distribution is approximately Gaussian (upper panel), the corresponding importance weights exhibit pronounced outliers (lower panel), which substantially reduce the ESS.  

To address this, we introduce an ESS-guided outlier-removal procedure. The replicas are first sorted in descending order of their weights $W_k^{(0)}$, and the largest-weight replicas are removed sequentially. After each removal, the ESS is recomputed. We observe that the ESS initially increases as extreme outliers are removed, reflecting the suppression of spurious concentration of weights. However, beyond a certain point, further removal leads to a decrease in ESS, indicating that genuinely important configurations are being discarded. We therefore stop the removal procedure at the point where the ESS is maximized. This procedure should therefore be regarded as a practical heuristic rather than a formally unbiased operation, since it modifies the weighted population used for initialization. We evaluate the sensitivity of the energy estimate to the number of removed samples using an independent comparison, as discussed in Sec.~\ref{sec:single-instance}.

After outlier removal, the number of remaining replicas is reduced from the initial population size $R$. Nevertheless, the resulting ESS is significantly increased, yielding a more reliable approximation of the target distribution. In some cases, however, even after outlier removal, the ESS remains insufficient. This observation motivates an adaptive selection of the initialization temperature, which we describe in the next subsection.  



\subsection{Adaptive selection of initialization temperature}
\label{sec:adaptive_t_set}
While the ESS-based outlier removal reduces the effect of extreme weights, it does not fully resolve the discrepancy between the TN-based initialization distribution and the canonical distribution when the initialization temperature is too low. To examine this effect, we study the ESS obtained from TN-based initialization at several temperatures. 

\begin{figure}
    \centering
 \includegraphics[width=\linewidth]{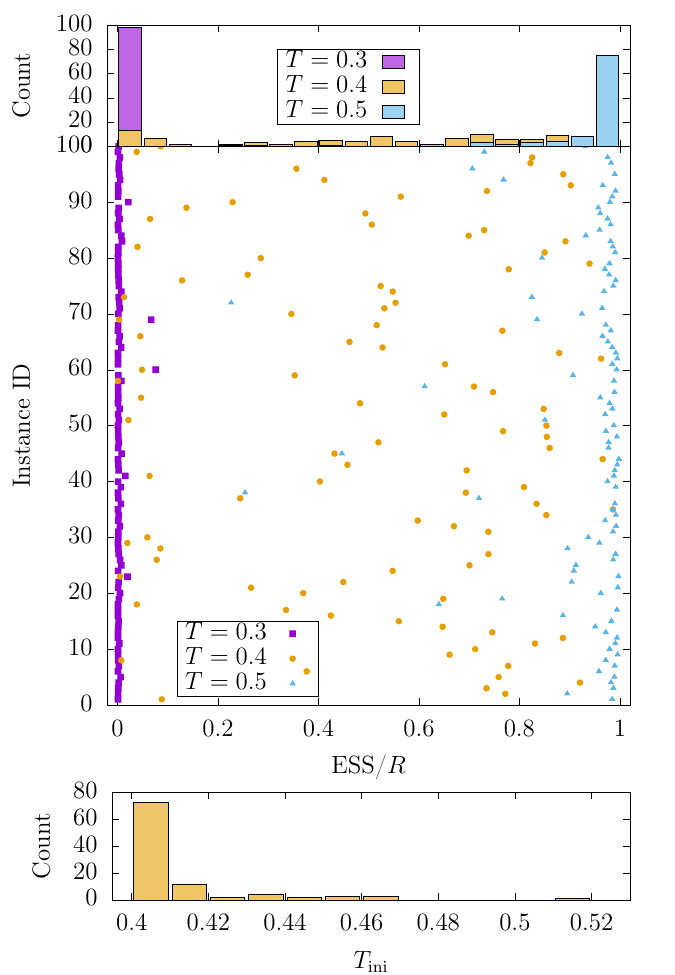}
    \caption{ESS ratios for 100 random instances of the two-dimensional EA $\pm J$ Ising model with $L=128$. Top: Histograms of $\mathrm{ESS}/R$ at fixed temperatures $T=0.3$, $0.4$ and $0.5$. Middle: Instance-by-instance values of $\mathrm{ESS}/R$ at the same temperatures. Bottom: Distribution of initialization temperatures selected by the adaptive procedure starting from $T=0.4$, which increases the temperature only when the ESS remains insufficient after outlier removal. The precise model definition and units are given later.
    }
    \label{fig:ess_analysis}
\end{figure}

Fig.~\ref{fig:ess_analysis} shows the distribution of ESS ratios for different disorder instances at three relatively low temperatures. At $T=0.5$, most instances yield large ESS ratios, indicating good overlap between $P_\mathrm{TN}$ and the canonical distribution at the corresponding temperature. At $T=0.4$, the ESS varies significantly across instances, indicating that this temperature lies near the boundary where TN-based initialization remains reliable. However, in many cases, the ESS can be recovered to an acceptable level by the ESS-based outlier removal described above, suggesting that the discrepancy is often caused by a small number of highly weighted samples. At $T=0.3$, by contrast, the ESS is substantially reduced in nearly all instances. These results demonstrate that TN-based sampling becomes unreliable at such low temperatures due to numerical instabilities in the TN contraction. 

Based on these observations, we adopt the following adaptive strategy. Starting from $T=0.4$ as the default candidate initial temperature, we first apply the ESS-based outlier removal to the $R$ replicas generated by TN-based sampling. If the resulting ESS falls below a prescribed fraction of $R$, e.g., $0.8R$ in this work, we increase the temperature (equivalently, decrease $\beta_0$) and repeat the procedure until a satisfactory ESS is obtained. In this way, instances for which outlier removal already works well can be initialized at relatively low temperatures, whereas instances that would otherwise require excessive removal are initialized at slightly higher temperatures. The resulting distribution of selected initialization temperatures is shown in the bottom panel of Fig.~\ref{fig:ess_analysis}. Most instances are initialized near $T\approx 0.40$, while a smaller fraction requires somewhat higher temperatures, indicating a moderate but non-negligible instance dependence in the temperature range where TN-based initialization remains reliable. 

\section{Results}\label{sec:results}
\subsection{Model: Two-dimensional $\pm J$ Ising spin glass}
We first specify the model used in the numerical experiments. In this study, we consider the two-dimensional $\pm J$ Edwards--Anderson (EA) Ising spin-glass model on a square lattice of linear size $L$, with $N=L^2$ spins. For a configuration $\vb*\sigma=(\sigma_1,\cdots,\sigma_N)$ where $\sigma_i\in\{-1,+1\}$ for each lattice site $i$, the Hamiltonian is 
\begin{equation}
  \mathcal{H}(\vb*\sigma)\;=\;-\sum_{\langle ij\rangle} J_{ij}\,\sigma_i \sigma_j,
  \label{eq:ising-ham}
\end{equation}
where the sum runs over nearest-neighbor bonds $\langle ij\rangle$ on the lattice. Unless otherwise stated, we use free boundary conditions, which are convenient for the TN construction. 

For this particular two-dimensional zero-field Ising model, specialized exact methods based on Pfaffian calculations are available. These include polynomial-time algorithms for evaluating the partition function~\cite{Saul1994641, Galluccio} and an exact finite-temperature algorithm for sampling spin configurations from the Boltzmann distribution~\cite{ThomasMiddleton2009}. In the following, we use the Pfaffian evaluation of the partition function to obtain exact energy benchmarks for comparison with the sampling results. 

The couplings $\{J_{ij}\}$ are quenched i.i.d.\ random variables taking values $\pm J$ according to  
\begin{equation}
  P(J_{ij})=p \delta\left(J_{ij},+J\right) + (1-p)\delta\left(J_{ij},-J\right),
\end{equation}
where $p$ denotes the probability for a ferromagnetic bond and $J$ is used as the unit of temperature $T$ in the following. In this work, we focus on the symmetric case $p=1/2$, in which ferromagnetic and antiferromagnetic bonds are equally distributed, and disorder is maximal. 
Thermal averages at inverse temperature $\beta$ are denoted by $\langle \cdots \rangle$, taken with respect to the canonical distribution, while disorder averages over the bond distribution are denoted by $[\cdots]_{\mathrm{av}}$. 

For the two-dimensional $\pm J$ EA spin glass, it is widely believed that no finite-temperature spin-glass phase transition occurs~\cite{Bray_1984, BhattYoung}. At the same time, a nonzero residual entropy is expected at zero temperature, reflecting the macroscopic degeneracy of the ground state. These features make the disorder-averaged entropy and its zero-temperature extrapolation natural quantities to examine.

\subsection{Tensor-network-based initialization}
\begin{figure}
    \centering
    \includegraphics[width=0.75\linewidth]{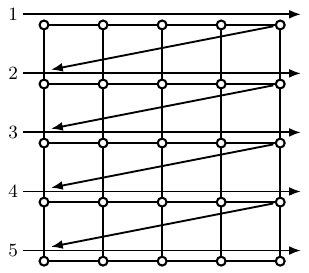}
    \caption{Zig-zag ordering of spins used in the autoregressive factorization on the two-dimensional square lattice. }
    \label{fig:zig-zag}
\end{figure}

To initialize the population in TNPA, we generate configurations by TN-based autoregressive sampling at an instance-dependent initialization temperature. The adaptive selection of this temperature based on the ESS was described in Sec.~\ref{sec:adaptive_t_set}. Here, we summarize the TN construction used for the initialization. All TN contractions in this initialization procedure are performed with bond dimension $\chi=16$. 

The ordering of spins in Eq.~(\ref{eqn:conditional}) is not unique. In this work, we adopt the zig-zag ordering on the two-dimensional square lattice, shown in Fig.~\ref{fig:zig-zag}, as it is a natural choice for the sequential TN contraction. Using the approximate conditional probabilities in Eq.~(\ref{eqn:conditional_partition}), we generate each configuration sequentially according to this ordering. At finite bond dimension $\chi$, the resulting TN distribution $P_{\mathrm{TN}}$ generally depends on the ordering, and even within the class of zig-zag orderings, there remains some freedom, such as the choice of the starting corner. We examined several such alternative orderings over a range of temperatures. At lower temperatures, where numerical instabilities begin to appear in the TN contraction, the TN-generated distributions exhibit a pronounced ordering dependence. At $T=0.4$, the default candidate initialization temperature used in this work, the physical observables obtained with the tested orderings are instead consistent within statistical uncertainty.  

As shown in Fig.~\ref{fig:ess_analysis}, the ESS of the resulting TN-based samples depends strongly on both the disorder instance and the temperature. Therefore, we choose the initialization temperature adaptively for each disorder instance according to the ESS criterion introduced in Sec.~\ref{sec:adaptive_t_set}.  This procedure yields an instance-dependent initialization temperature at which TN-based sampling provides a sufficiently large ESS without excessive outlier removal. 

\subsection{MCMC updates and temperature schedule}
For the MCMC updates within PA, we employ isoenergetic cluster Monte Carlo (ICM)~\cite{Houdayer, ICM}. In ICM, cluster moves are performed on pairs of replicas by flipping connected regions with pair overlap equal to $+1$, which preserves both the mutual overlap structure and the total energy of the replica pair. Its combination with PA, therefore, requires careful handling of the resampling step. In particular, when ICM is applied to pairs of replicas, resampling must also be carried out pairwise in order to maintain consistency with the pairwise update procedure. Further details are given in Appendix~\ref{sec:combinationICMPA}

Since ICM alone does not guarantee irreducibility of the Markov chain, we combine it with Gibbs sampling. In this work, one Monte Carlo step (MCS), or one sweep, is defined as one full Gibbs-sampling update over all spins. In TNPA and conventional PA, each temperature step consists of 500 sweeps, followed by one ICM update, followed by another 500 sweeps.  In EXMC, the update block between adjacent replica-exchange attempts consists of 5 sweeps, one ICM update, and 5 additional sweeps.    

Throughout the TNPA calculations, we adopt a systematic resampling scheme and perform resampling at every temperature step, since this choice provides the best diversity retention, as measured by the family entropy, among the schemes we tested; see Appendix~\ref{sec:resampling_family_entropy}. The temperature schedule for TNPA is determined adaptively rather than by using a fixed temperature increment. Specifically, when moving from one temperature to the next, the next temperature is chosen so that the ESS after the weight update satisfies $\mathrm{ESS}\approx 0.99 R$. In other words, the temperature step is adjusted to keep weight degeneracy weak at each stage. 

Several adaptive temperature-scheduling schemes have been proposed for PA and related Sequential Monte Carlo methods, including criteria based on the ESS~\cite{Jasra_2011, Zhou02072016}, the overlap between neighboring energy distributions~\cite{BARASH2017341, Christiansen_2019}, and the culling fraction and energy variance~\cite{Amery-Machta-2018}. Here, we employ the ESS-based criterion described above. Although formulated differently, these criteria are closely related because they control changes in the statistical weights between neighboring temperatures, which are governed by energy fluctuations. In particular, for sufficiently small temperature increments, maintaining a fixed ESS ratio gives $\Delta\beta,\sigma_E(\beta)\simeq \mathrm{const.}$. 

In general, frequent resampling can reduce population diversity because highly weighted replicas are repeatedly selected. In the present approach, however, this loss of diversity is efficiently compensated by the ICM updates, which provide large-scale nonlocal rearrangements of the configurations while preserving the relevant overlap structure. After each resampling step, the replica pairing is refreshed before the next ICM update.  

When disorder-averaged quantities are computed across multiple instances, a common temperature schedule must be used for all instances. In this case, we first determine the adaptive temperature schedule for the instance with the highest initialization temperature. We then use this same temperature schedule for all other disorder instances in the disorder-averaged calculations. 

\subsection{Single-instance comparison with existing methods}
\label{sec:single-instance}
We begin by examining a single disorder instance to evaluate the equilibration achieved by TNPA. For this purpose, we compare the temperature dependence of the energy density and the overlap distribution $P_J(q)$, which is a central observable in spin-glass studies, with the corresponding results obtained using existing methods. 

We first study the internal energy density $e(T)=\langle \mathcal{H}(\vb*\sigma)\rangle/N$ as a function of temperature $T$ using three sampling methods, TNPA, conventional PA and EXMC, together with an exact Pfaffian calculation. In TNPA, the population is initialized by TN-based sampling at $T=0.4$ with population size $R=10^3$, and PA is then run down to $T=0.2$. By contrast, conventional PA is started from the high-temperature limit and cooled to $T=0.2$ with the same population size, $R=10^3$. Both TNPA and conventional PA use $130$ temperature points and terminate at $\beta=5$. For TNPA, the temperature sequence is determined adaptively from $\beta=2.5$ using the ESS-based criterion, whereas conventional PA uses equally spaced inverse temperatures starting from $\beta=0.2$. The resulting temperature schedules are shown in Appendix~\ref{sec:temperature_schedule}.  In this comparative numerical experiment, we adjusted parameters such as the number of MCMC sweeps per temperature so that the total computational cost of PA and TNPA was approximately the same, thereby enabling a fair comparison. 

For reference, we also include the exchange Monte Carlo method~\cite{Geyer1991, HukushimaNemoto} (EXMC), also known as parallel tempering~\cite{Marinari1998}. EXMC maintains replicas on a fixed temperature ladder and exchanges configurations between neighboring temperatures according to a Metropolis-type acceptance criterion. As a result, each replica performs a random walk in temperature space and can overcome low-temperature free-energy barriers by transitioning to higher temperatures. By contrast, PA evolves an entire population sequentially along an annealing schedule, using reweighting and resampling at each temperature step. Thus, EXMC accelerates equilibration through replica exchanges across temperatures, whereas PA evolves the population through sequential reweighting and resampling.
   
For the energy comparison, we additionally employ an exact Pfaffian calculation. For the planar Ising model with free-boundary conditions considered here, the partition function can be evaluated exactly using the Pfaffian formulation~\cite{Saul1994641, Galluccio}, from which we obtain the numerically exact internal energy. In contrast to EXMC, the Pfaffian calculations in this study provide an independent validation distinct from the sampling method. 
\begin{figure}
    \centering
    \includegraphics[width=\linewidth]{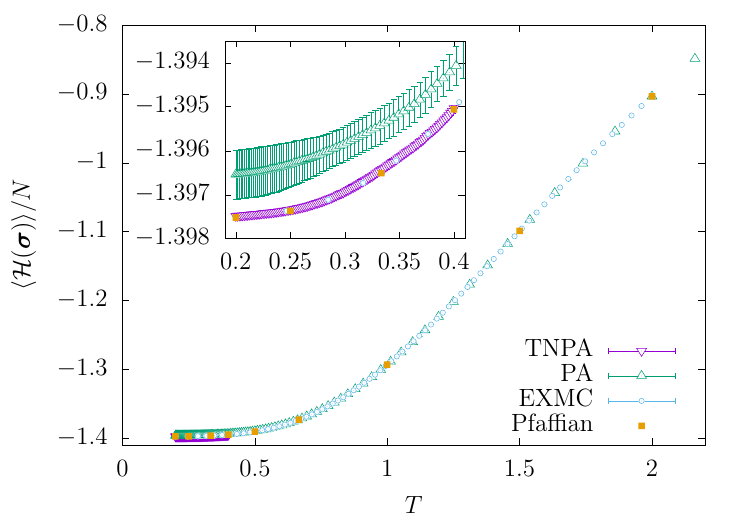}
    \caption{Temperature dependence of the energy density of the two-dimensional EA Ising model for a typical disorder instance with $L=128$. The inset shows an enlarged view of the low-temperature regime, $0.2\leq T\leq 0.4$. 
    The PA results are averaged over 10 independent runs, with the error bars denoting the corresponding standard errors. The TNPA and EXMC results are obtained from single runs. The filled squares show the numerically exact values obtained from the Pfaffian calculation.  
    }
    \label{fig:energy}
\end{figure}

As shown in Fig.~\ref{fig:energy}, the energy densities obtained by TNPA and EXMC agree with the exact Pfaffian values within statistical uncertainties over the temperature range examined. At high temperatures, conventional PA also agree with the exact values, indicating that it remains well equilibrated in this regime. In the enlarged low-temperature view, however, the energy obtained by conventional PA remains systematically higher than the exact value, indicating insufficient equilibration. Although PA yields correct equilibrium sampling in the infinite-population limit $R\to\infty$ regardless of the cooling rate, in practice the diversity of a finite population can decrease during annealing from high to low temperatures, potentially leading to insufficient sampling in the low-temperature regime.  

As an additional robustness check, we varied the number of removed samples for the same representative instance. The reweighted energy density remained stable over a broad range of removal thresholds, including the value determined by maximizing the ESS, and remained consistent with the exact Pfaffian value within statistical uncertainty. We therefore found no statistically detectable removal-induced bias in the energy density in this test, although this does not establish that the procedure is unbiased in general. 

In contrast, TNPA generates its initial population at the finite temperature $T=0.4$ by TN-based sampling and avoids the long annealing process from the high-temperature limit. This makes it easier to maintain population diversity even in subsequent low-temperature simulations and improves equilibration. Even when PA and TNPA are compared at approximately the same computational cost, Fig.~\ref{fig:energy} shows that a systematic deviation in the conventional PA energy is already observed at the TNPA-initialized temperature, $T=0.4$. In other words, under the same computational resources, the systematic energy deviation observed for conventional PA at $T=0.4$ is consistent with the population having lost diversity during annealing from the high-temperature limit, thereby insufficiently exploring the low-temperature region.    

\begin{figure}
    \centering
   \includegraphics[width=\linewidth]{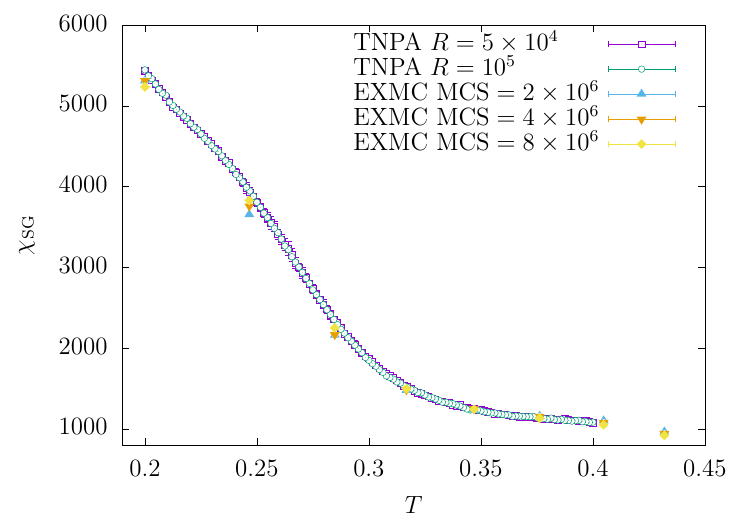}
    \caption{Temperature dependence of the spin-glass susceptibility of the two-dimensional EA Ising model for a typical disorder instance with $L=128$. The population sizes used in TNPA are $5\times 10^4$ and $10^5$. For EXMC, the total number of MC steps are $2\times 10^7$, $4\times 10^7$, and $8\times 10^7$. }
    \label{fig:sg_susceptibility}
\end{figure}

\begin{figure}
    \centering
    \includegraphics[width=\linewidth]{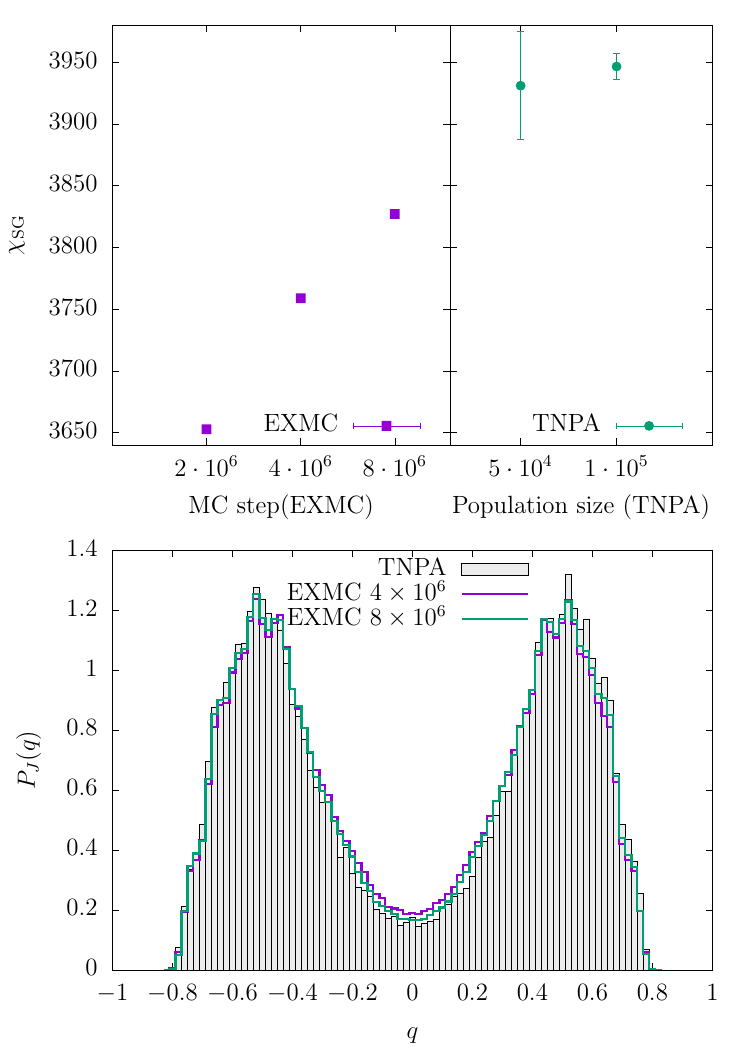}
    \caption{Top: Dependence of the spin-glass susceptibility on the number of MC steps in EXMC and on the population size $R$ in TNPA at the lowest temperature studied, $T=0.247$. Bottom: Overlap distribution $P_J(q)$ at the same temperature for the same disorder realization, obtained by EXMC and TNPA. }
    \label{fig:pofq}
\end{figure}

Next, we compare the spin-glass susceptibility $\chi_\mathrm{SG}$ and the overlap distribution $P_J(q)$. 
For two independent replicas, $\vb*{\sigma}^{(1)}$ and $\vb*{\sigma}^{(2)}$, drawn at the
same temperature for the same disorder realization, the overlap distribution is defined as 
\begin{equation}
  P_J(q) \;=\; \left\langle\delta\left( Nq, \ \sum_{i=1}^{N} \sigma_i^{(1)} \sigma_i^{(2)}\right)\right\rangle, 
  \label{eq:overlap}
\end{equation}
where $\delta(\cdot,\cdot)$ denotes the Kronecker delta. In TNPA, this quantity is estimated by selecting two replicas with replacement from the population according to their weights. After resampling, the replica weights are reset to unity; therefore, the same quantity can equivalently be estimated by uniformly sampling two replicas from the resampled population. By contrast, in EXMC, $P_J(q)$ is evaluated from two independent Markov chains sampled at the same temperature. The spin–glass susceptibility is given by the second moment of $P_J(q)$, 
\begin{equation}
    \chi_{\mathrm{SG}} \;=\; N\,\big\langle q^2 \big\rangle. 
    \label{eq:chi_sg_from_q}
\end{equation}

In TNPA, we use population sizes $R=5\times 10^4$ and $10^5$; in EXMC, the total number of MC steps is increased from $2\times 10^7$ to $4\times 10^7$ and $8\times 10^7$. As shown in Fig.~\ref{fig:sg_susceptibility}, $\chi_\mathrm{SG}(T)$ varies smoothly over the temperature range $0.2\leq T\leq 0.4$, and the TNPA results for the two different population sizes nearly coincide. Thus, at least for this disorder instance, TNPA exhibits weak population-size dependence and yields stable estimates. 

By contrast, the values of $\chi_\mathrm{SG}$ obtained by EXMC increase at low temperatures as the total number of MC steps increases. The top panel of Fig.~\ref{fig:pofq} shows the MC-step dependence of $\chi_\mathrm{SG}$ at the lowest temperature studied for EXMC, along with the population-size dependence of TNPA. While the TNPA estimates are consistent within statistical uncertainty, the EXMC estimates shift toward larger values as the MC step increases. This indicates that EXMC has not yet converged at this temperature. At the same time, the EXMC data appear to be approaching the TNPA estimate, supporting the interpretation that TNPA is already close to the equilibrium value.   

Furthermore, to examine this trend in more detail, we compare the overlap distribution $P_J(q)$ at the same temperature (see the bottom panel of Fig.~\ref{fig:pofq}). Although the difference is small, the EXMC results have slightly more weight near $q\simeq 0$ than the TNPA results. This residual central weight is consistent with the smaller value of $\chi_\mathrm{SG}$, which is given by the second moment of $P_J(q)$. As the number of MC steps increases, the EXMC distribution gradually approaches that obtained by TNPA. Nevertheless, even within the longest run considered here, a visible excess weight remains near the center, indicating that convergence to the low-temperature equilibrium distribution is still incomplete. 

This comparison is more comprehensive than one based solely on energy density. Indeed, the energy estimates obtained by TNPA and EXMC are similar, indicating that both methods provide comparable estimates of this single-replica thermodynamic observable. By contrast, $P_J(q)$, which is constructed from two independent replicas, examines the structure of low-lying spin configurations and their relative statistical weights in greater detail. A plausible interpretation is that, whereas EXMC explores new states through round trips to higher temperatures, TNPA achieves broader coverage of low-temperature states through its large population of replicas. At least for the present system size, this suggests that TNPA maintains sufficient diversity despite repeated resampling. 

\subsection{Disorder-averaged entropy and residual entropy}
\label{sec:disorder_averaged}
We now turn to physical quantities averaged over disorder realizations. Our main focus here is the entropy and its extrapolation to zero temperature, from which the residual entropy is estimated. In conventional MCMC-based approaches, entropy is typically obtained only indirectly, for example through thermodynamic integration, or more directly by estimating the density of states using multicanonical~\cite{multi-canonical} and flat-histogram~\cite{Wang1999} methods. In the present approach, by contrast, we exploit the free-energy estimator naturally available in PA.  

In PA, the free-energy difference between successive temperatures is estimated from the normalization factor of the reweighting step, before resampling is applied. Specifically, for a temperature step from $\beta_i$ to $\beta_{i+1}$, we define 
\begin{equation}
    \overline{W}^{(i+1)} = \frac{1}{R}\sum_{k=1}^R W_k^{(i+1)}.
\end{equation}
The free energy then satisfies 
\begin{equation}
-\beta_{i+1} F(\beta_{i+1}) = -\beta_i F(\beta_i) + \ln \overline{W}^{(i+1)}.
\end{equation}
Resampling is performed only after this normalization factor has been evaluated. Note that resampling only restores an unweighted population and does not change the free-energy estimate. 
Using this free-energy estimate together with the internal energy, we evaluate the entropy for each disorder realization $J$ as 
\begin{equation}
    S_J(T,L) = \frac{\langle  \mathcal{H}\rangle_J-F_J(T)}{T}.
\end{equation}
For a finite system, both $S_J(T,L)$ and its zero-temperature limit depend on the disorder realization, indicating the instance dependence of the ground-state degeneracy. We subsequently average $S_J(T,L)$ over disorder realizations, as described below. 

In TNPA, the PA run starts from an instance-dependent initialization temperature rather than from the high-temperature limit. Therefore, to obtain the entropy on an absolute scale, we also evaluate the entropy difference between the infinite-temperature limit and the corresponding initialization temperature from the TN-based weight factors used in the initialization step. Since the entropy at infinite temperature is exactly $N\ln 2$, this allows us to determine the temperature dependence of the entropy over the entire range and, in particular, to estimate the residual entropy by extrapolating to zero temperature. 

\begin{figure}
    \centering
    \includegraphics[width=\linewidth]{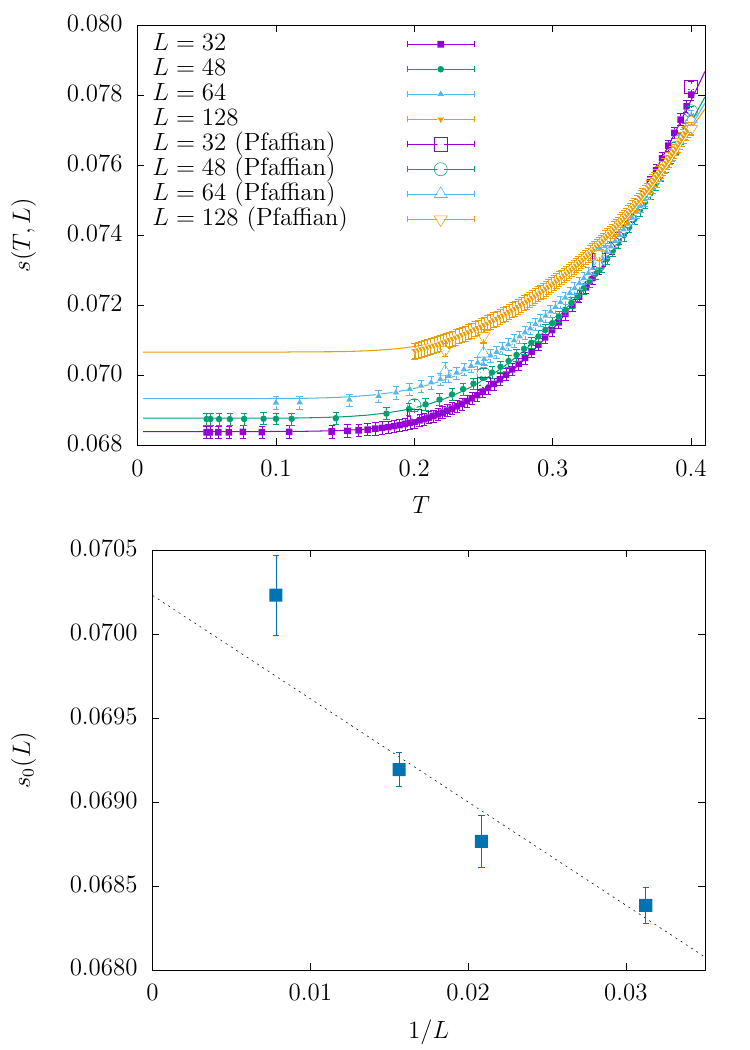}
    \caption{Top: Temperature dependence of the disorder-averaged entropy density of the two-dimensional EA spin glass for system sizes $L=32$, $48$, $64$, and $128$. The filled and open symbols represent the estimates obtained using TNPA and the Pfaffian method, respectively. The error bars denote standard errors over independently generated disorder realizations, and the solid curves show fits for $T\leq 0.3$, which are used to extrapolate the residual entropy density $s_0(L)$. Bottom: System-size dependence of the extrapolated residual entropy density $s_0(L)$. The error bars represent the uncertainties obtained from the low-temperature extrapolation, and the dotted line shows a linear fit in $1/L$. }
    \label{fig:entropy}
\end{figure}

We first show the temperature dependence of the disorder-averaged entropy density $s(T, L)=[S_J(T, L)]_\mathrm{av}/N$ for finite systems in the top panel of Fig.~\ref{fig:entropy}. The system sizes examined in this work are $L=32$, $48$, $64$ and $128$, for which the disorder averages are taken over $N_\mathrm{inst}=2000$, $1000$, $500$ and $100$ instances, respectively. The population size is fixed to $R=10^5$ for $L=32$, $48$, and $64$, while for $L=128$ we use $R=10^5$ and $2\times 10^5$. For comparison, we also estimate the disorder-averaged entropy density using the Pfaffian method. For each system size, the Pfaffian average is taken over an independent set of at least 100 disorder realizations. Within the temperature range in which the Pfaffian calculation remains numerically stable, the Pfaffian and TNPA estimates are consistent within one standard deviation. At lower temperatures, the Pfaffian calculation becomes numerically unstable, particularly for larger systems. 

Based on the TNPA results, we estimate the residual entropy density $s_0(L)=\lim_{T\to 0}s(T, L)$, by fitting the low-temperature behavior. In the fitting formula based on a low-temperature expansion, we denote the first excitation gap by $J$. In this context, the form of the leading low-temperature correction has long been discussed in relation to the low-temperature specific heat. Even under periodic boundary conditions, there is a subtle issue that contributions with energy scale $2J$, although not usually regarded as the leading elementary excitations, can nevertheless appear effectively in finite-size analyses~\cite{WangSwendsen_1988, Lukic_2004}. In the present work, the use of free boundary conditions in the TN calculations may introduce additional subtleties. We do not pursue this issue in detail here; instead, we perform the extrapolation using a fitting form that includes terms up to second order in the low-temperature expansion associated with the $J$ excitation, which may be related to the use of free boundary conditions. 

Next, we extrapolate the residual entropy density to the thermodynamic limit using the finite-size estimates obtained in this way. As shown in the bottom panel of Fig.~\ref{fig:entropy},  a linear fit in $1/L$ gives 
\begin{equation}
    s_0 = 0.07023(34)
\end{equation}
This value is consistent with previous estimates listed in Table~\ref{tab:residual_entropy}. Because only four system sizes are available and subleading finite-size corrections cannot be excluded, the quoted uncertainty should be interpreted as the statistical uncertainty within the assumed linear extrapolation. Although the present estimate does not yet demonstrate a clear quantitative advantage over existing approaches, it is noteworthy that TNPA reaches system sizes comparable to those of the largest Monte Carlo-based studies and yields an estimate near to the lower end of the reported range. These results imply that TNPA can provide lower-temperature entropy estimates for large systems that are suitable for a finite size analysis of the residual entropy.

\begin{table}[]
\caption{\label{tab:residual_entropy}%
Previous estimates of the residual entropy density $s_0$ of the two-dimensional $\pm J$ EA Ising spin glass. The table includes transfer-matrix methods, exact ground-state search algorithms such as branch-and-bound, combinatorial methods, exact partition-function approaches, and Monte Carlo-based approaches. $N_\mathrm{max}$ denotes the largest system size treated in each work.   
}
\begin{ruledtabular}
\begin{tabular}{lllll}
\textrm{Ref.}&
\textrm{Year}&
\textrm{Method}&
$N_\mathrm{max}$&
$s_0$\\
\colrule
\cite{Cheung_1983} & 1983 &Transfer-matrix method & $11\times\infty$ & 0.0701(5) \\
\cite{WangSwendsen_1988}& 1988 & Replica MC &  $128^2$
& 0.071(7)\\
\cite{Saul1994641} & 1994 & Exact ground-state search & $36^2$ & 0.075(2) \\
\cite{Blackman_1998} & 1998 & Combinatorial method & $256^2$ &  0.0709(4) \\ 
\cite{ZHAN2000239} & 2000 & Flat-histogram MC & $36^2$ & 0.0709(60) \\ 
\cite{Hartmann_2000}  & 2000 & Exact ground-state search & $20^2$ & 0.078(5)  \\
\cite{Lukic_2004} & 2004 & Exact partition function & $50^2$ & 0.0714(2) \\
\cite{Perez-Morelo_2012} & 2012 & Parallel tempering & $20^2$ & 0.0714(9)
\end{tabular}
\end{ruledtabular}
\end{table}

\section{Summary and discussion}
\label{sec:summary}
In this work, we proposed TNPA, which combines TN methods with PA, and applied it to low-temperature simulations of the two-dimensional $\pm J$ EA Ising spin glass. The proposed method integrates TN-based sampling of nearly independent initial configurations, large-scale parallel annealing, and cluster updates using ICM. This combination enables stable equilibration even in low-temperature regimes that are difficult to reach by TN-based methods alone, and in challenging systems for which local-update MCMC methods exhibit extremely slow relaxation. 

The central idea of TNPA is that TN-based sampling and PA play complementary roles, so that the weaknesses of each method are compensated by the strengths of the other. From the PA perspective, TN provides initial samples that are already close to the target distribution, thereby reducing the need for a long annealing process from the high-temperature limit and suppressing the loss of population diversity. From the TN perspective, nearly independent samples are generated within a temperature range where TN contractions remain relatively stable, and the subsequent low-temperature regime is then handled by PA. This enables stable evaluation of physical observables even in regimes where direct TN calculations become numerically unstable. Furthermore, once sampling is achieved, observables can be evaluated directly by sample averages, without introducing observable-specific impurity tensors or relying on numerical differentiation of the free energy. In this sense, TNPA provides a flexible route to quantities that are cumbersome to access in conventional TN-based approaches, including spin-glass observables such as the overlap distribution and the spin-glass susceptibility. 

Before concluding, we briefly discuss several points that are worth further consideration. 
First, we discuss the computational cost of TNPA, in particular that associated with the TN-based initialization. The present implementation is restricted to two-dimensional systems, and we do not yet have sufficiently controlled results for the three-dimensional spin-glass models. In the present study, a large population size of order $2\times 10^5$ was required to achieve equilibration for difficult disorder instances at low temperatures. A direct extension to three-dimensional spin-glass models or to larger system sizes would therefore require further methodological development and substantially greater computational cost. One possible way to improve efficiency is to perform TN-based sampling not on the full system but on local regions or blocks. For example, a promising direction for three-dimensional extensions is to use two-dimensional planes as the basic sampling blocks~\cite{chen2025_3d}. While increasing the block size enables more collective updates over larger regions, it also introduces a trade-off with TN contraction accuracy, or equivalently with the bond dimension required to maintain the desired accuracy. Developing criteria for choosing the optimal block size and adaptive update strategies remains an important challenge for future work~\cite{Gantan_2024, chen2025_3d}. 

Next, we discuss the relation to machine-learning-based sampling methods. In recent years, machine-learning-model-based approaches, including autoregressive models, have attracted considerable attention in MCMC sampling for spin-glass models~\cite{Wu_2019, Del_Bono_2025, Bono_PRE_2025, McNaughton_2020, Wu2021, Inack2022, Ciarella_2023}. At the same time, while some studies have reported encouraging improvements in sampling performance, others have pointed out that such methods do not necessarily remain effective on hard instances. In this respect, TN-based approaches have the distinct advantage of not requiring a training stage and instead provide a direct approximation derived from the underlying physical model. 

From the perspective of the present work, it may also be useful to monitor ESS-like quantities and the influence of rare, overly weighted samples in such machine-learning-based approaches. This may be particularly relevant in iterative self-training schemes, where a learning model is refined using samples generated by its current approximation. In such a setting, outliers produced by an inaccurate intermediate model may affect the refinement process, and the ESS-based diagnostics and outlier-removal procedure introduced here may help to reduce this effect. More broadly, there may be a structural relation between the distributions learned by machine-learning models and those approximated by TNs under a finite bond-dimension constraint. Investigating this relation may deepen our understanding of both approaches and suggest new hybrid algorithms. 

Finally, we discuss the instance dependence of the TN approximation accuracy. As shown in this work, the quality of the TN approximation depends strongly on the disorder instance. In particular, we introduced an adaptive procedure to select the initialization temperature that maintains a sufficiently large ESS, and found that the resulting initialization temperature varies from instance to instance. From this viewpoint, instances requiring higher initialization temperatures may be regarded as more difficult for the TN method. 

However, it remains unclear what fundamentally determines the difficulty of a disorder instance for TN approximation. In homogeneous interacting systems, it is well known that the accuracy of TN methods is strongly affected by the correlation length of the system~\cite{Tagliacozzo_2008}, and that the approximation tends to become less accurate near a second-order phase transition, where the correlation length diverges. In disordered systems, such as random spin models, we have not found any clear correlation between the required initialization temperature and simple indicators, such as correlation length or frustration strength. This suggests that the difficulty for TN may arise from a more intricate combination of factors. Identifying quantitative indicators of such instance-dependent difficulty would be an important direction for future work, as it could lead to improved TN approximations and more efficient sampling strategies. 

\begin{acknowledgments}
The authors would like to express their special thanks to Yamato Arai for his continuous and valuable discussions. 
This work was supported by JSPS KAKENHI Grant Number 23H01095 and JST Grant Number JPMJPF2221.
The computations in this work were partially performed using the facilities of the Supercomputer Center, the Institute for Solid State Physics, the University of Tokyo (ISSPkyodo-SC-2024-Cb-0041, 2025-Ca-0081).
\end{acknowledgments}

\section*{data availability}
The numerical data underlying the results and figures presented in this paper are openly available in Zenodo at \url{https://doi.org/10.5281/zenodo.23181327}. 

\appendix
\section{Combining Isoenergetic cluster Monte Carlo with population annealing}
\label{sec:combinationICMPA}
When isoenergetic cluster Monte Carlo (ICM) is combined with population annealing (PA), particular care is required in the resampling step. In standard implementations, replica pairs are typically selected sequentially at random, and cluster updates are applied to the selected pairs. To examine the consistency of this combination, we consider here a simplified setting in which the inverse temperature is decreased with a constant step size, while resampling is triggered whenever the ESS ratio falls below a prescribed threshold. Under such a sequential random-pairing scheme, we found that the estimated equilibrium energy depends on the chosen ESS threshold. Since equilibrium expectation values should not depend on this auxiliary parameter, this behavior indicates that the sampling procedure is not correctly reproducing the equilibrium distribution.     

A possible origin of this problem is that ICM introduces correlations between paired replicas. In particular, because the total energy of a replica pair is preserved, a cluster flip that increases the energy of one replica necessarily decreases that of the other. By contrast, ordinary resampling procedures treat replicas as independent. This inconsistency can lead to distorted effective population weights and biased equilibrium estimates. To avoid it, we employ a pairwise resampling scheme. At each temperature step, we first generate a uniformly random perfect matching of all $R$ replicas, thereby forming $R/2$ fixed pairs. ICM updates are then applied to these fixed pairs, while the pairing is kept unchanged until the next resampling step.    

Resampling is performed at the pair level rather than at the level of individual replicas. For a pair $(a,b)$, we define the pair weight as $W_{ab} = W_a + W_b$ where $W_a$ and $W_b$ are the weights of the two replicas. Using systematic resampling, offspring numbers are then assigned according to these pair weights. After resampling, the weights of both members of each surviving pair are reset to unity, and before the next ICM update, the replicas are reshuffled to form a new random perfect matching. 

\begin{figure}
    \centering
    \includegraphics[width=\linewidth]{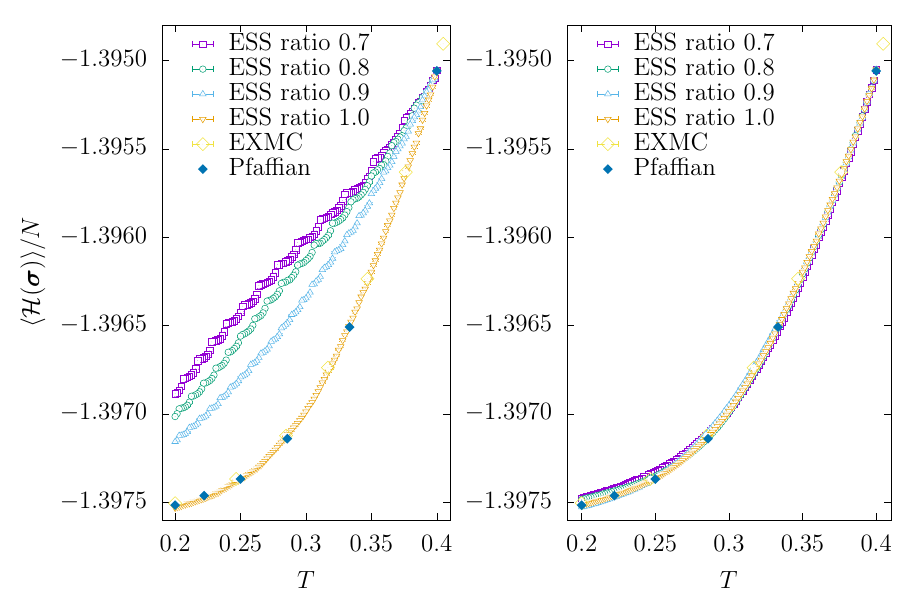}
    \caption{Temperature dependence of the energy density of the two-dimensional EA Ising spin-glass model with $L=128$ for a single disorder instance. The left and right panels show the results obtained using the sequential random-pairing scheme and the proposed pairwise resampling scheme, respectively. In each panel, curves for several ESS-ratio thresholds are shown. For comparison, the corresponding EXMC results and the numerically exact values obtained from the Pfaffian calculation are also included in both panels. }
    \label{fig:pairwise_resampling}
\end{figure}

Figure~\ref{fig:pairwise_resampling} demonstrates the effect of this modification. In the left panel, the sequential random-pairing scheme yields energy estimates that strongly depend on the ESS threshold and deviate from the exact Pfaffian values, indicating that the desired equilibrium is not achieved. By contrast, the proposed pairwise resampling scheme removes this threshold dependence almost completely, and the resulting energy estimates agree with the exact Pfaffian values over the temperature range examined. This shows that pairwise treatment is necessary to recover at least this basic requirement for equilibrium sampling, and strongly suggests that the resampling bias caused by ICM-induced correlations has been effectively suppressed.

\section{Resampling schemes and family entropy criterion}
\label{sec:resampling_family_entropy}
We compare four standard resampling schemes within PA: multinomial, systematic, stratified, and residual resampling: see, e.g., Ref.~\cite{ResamplingPA} for a detailed discussion of these schemes in the PA context and Refs.~\cite{ParticleFilter} for more general descriptions. Although these schemes differ in the distribution of offspring numbers at finite population size $R$, they share the same mean offspring numbers, i.e., the expected number of descendants of each replica is proportional to its normalized weight. As a result, they recover the same limiting population, as $R\to\infty$. Their finite-$R$ behavior, however, can differ substantially, in particular in the variance of estimators and in the loss of population diversity. 

To monitor the loss of genealogical diversity, we use the family entropy~\cite{Wang_2015, ResamplingPA} 
\begin{equation}
  S_{\mathrm f} \;=\; -\sum_{i:\,p_i>0} p_i \log p_i\, 
\end{equation}
where $p_i=n_i/R$ is the fraction of the population at a given temperature descending from ancestor $i$. Unlike the ESS, which measures weight degeneracy at a single temperature step, the family entropy probes the genealogical diversity of the population generated by repeated resampling.  

\begin{figure}
    \centering
    \includegraphics[width=\linewidth]{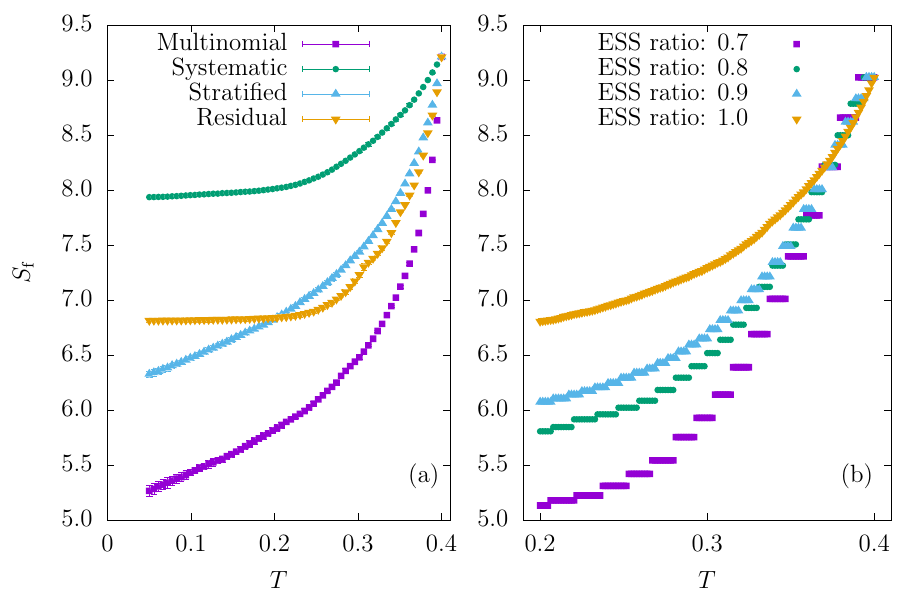}
    \caption{Temperature dependence of the family entropy during the PA run for the two-dimensional EA Ising spin-glass model. Left: Results for four different resampling schemes with $L=32$. Right: Results for different ESS thresholds for triggering resampling, with systematic resampling fixed, for $L=128$. }
    \label{fig:family_entropy}
\end{figure}

Figure~\ref{fig:family_entropy} shows the temperature dependence of $S_\mathrm{f}$ for the four resampling schemes. Among the schemes tested, systematic resampling exhibits the smallest reduction in family entropy, particularly in the low-temperature regime. We therefore adopt systematic resampling in all results reported in the main text. 

Keeping the resampling scheme fixed to systematic resampling, we further varied the ESS threshold used to trigger resampling. Here, a threshold of unity corresponds to resampling at every temperature step, whereas smaller thresholds trigger resampling only when the ESS falls below the prescribed value along the same ESS-guided adaptive temperature schedule. As shown in the right panel of Fig.~\ref{fig:family_entropy}, resampling at every step gives the best overall performance in our setting. We attribute this to frequent resampling, which suppresses weight degeneracy, while the accompanying MCMC updates efficiently restore diversity. 

\section{Adaptive temperature schedule}
\label{sec:temperature_schedule}
\begin{figure}
    \centering
    \includegraphics[width=\linewidth]{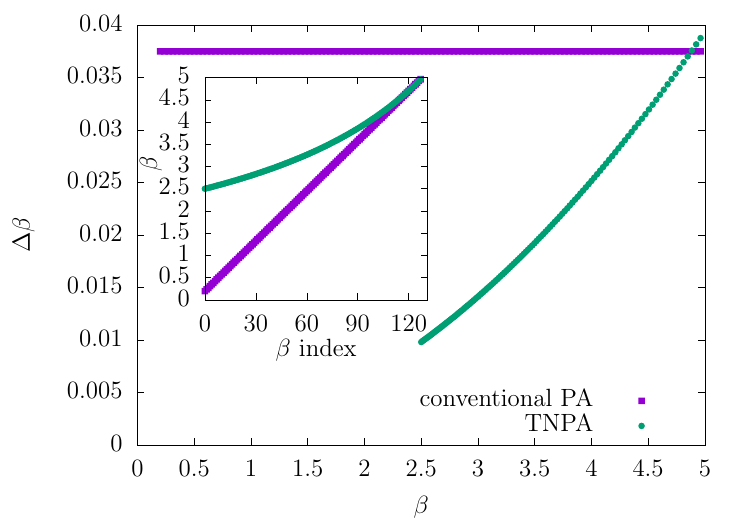}
    \caption{Inverse temperature schedules used for TNPA and conventional PA in the single-instance comparison. Main panel: Temperature increment $\Delta\beta_i=\beta_{i+1}-\beta_i$ as a function of $\beta_i$. TNPA stars from $\beta=2.5$, and each increment is determined adaptively so that $\mathrm{ESS}\simeq 0.99R$. Conventional PA starts at $\beta=0$, and its first inverse temperature is $\beta=0.2$, followed by equally spaced inverse temperature points. Both finite-temperature schedules contain $130$ points and terminate at $\beta=5$. Inset: Inverse temperature $\beta_i$ as a function of the temperature-step index. 
    }
    \label{fig:beta_schedule}
\end{figure}

Figure~\ref{fig:beta_schedule} shows the inverse-temperature schedules used for TNPA and conventional PA in the single-instance comparison presented in Sec.~\ref{sec:single-instance}. Both schedules contain $130$ temperature points and terminate at $\beta=5$. Conventional PA is initialized at $\beta=0$, corresponding to the infinite-temperature limit. Its first finite inverse temperature is $\beta=0.2$, and the subsequent temperature points are equally spaced. By contrast, TNPA starts from the TN initialization temperature $\beta_0=2.5$, corresponding to $T=0.4$, and determines each subsequent increment adaptively so that $\mathrm{ESS}/R\simeq 0.99$, where $R$ is the population size.  

As shown in the main panel of Fig.~\ref{fig:beta_schedule}, the temperature increment $\Delta\beta_i$ in TNPA increases toward lower temperatures. For sufficiently small increments, maintaining a fixed ESS ratio approximately implies $\Delta\beta_i\sigma_E(\beta_i)\simeq\mathrm{const.}$, where $\sigma_E(\beta_i)$ is the standard deviation of the energy. In the instance considered here, the energy distribution narrows as temperature decreases, allowing larger increments in $\beta$. The inset shows the resulting inverse temperature as a function of the temperature-step index. Since the same update sequence is performed at each temperature point, the computational cost of the PA stage is approximately proportional to the number of temperature points for a fixed population size.

\bibliography{patn_bib}

\end{document}